\documentclass[nofootinbib,showpacs,amsmath,amssymb,twocolumn]{revtex4}
\usepackage{graphicx}

\begin{document}
\title{\bf The strong form of the Levinson theorem for a distorted KP potential}
\author{Siamak S.
Gousheh,\footnote{Corresponding author, Phone: +98-(21)-299-02770,
Fax:       +98-(21)-224-31666, Electronic address:
ss-gousheh@sbu.ac.ir}
\email{ss-gousheh@sbu.ac.ir} Maryam
Taheri-Nejad} \email{m.taherinejad@sbu.ac.ir}
\affiliation{Department of Physics, Shahid Beheshti University G.
C., Evin, Tehran 19839, Iran}
\author{Mohammad R.
Fathollahi} \email{fathollahi@ee.kntu.ac.ir}
\affiliation{ Electrical Engineering Faculty, K.N. Toosi
University of Technology, 
Tehran,
16315, Iran}
\date{\today}
\begin{abstract}
We present a heuristic derivation of the strong form of the Levinson
theorem for one-dimensional quasi-periodic potentials. The
particular potential chosen is a distorted Kronig-Penney model. This
theorem relates the phase shifts of the states at each band edge to
the number of states crossing that edge, as the system evolves from
a simple periodic potential to a distorted one. By applying this
relationship to the two edges of each energy band, the modified
Levinson theorem for quasi-periodic potentials is derived. These two
theorems differ from the usual ones for isolated potentials in
non-relativistic and relativistic quantum mechanics by a crucial
alternating sign factor $(-1)^{s}$, where $s$ refers to the adjacent
gap or band index, as explained in the text. We also relate the
total number of bound states present in each energy gap due to the
distortion to the phase shifts at its edges. At the end we present
an overall relationship between all of the phase shifts at the band
edges and the total number of bound states present.
\end{abstract}

\maketitle

\section{Introduction}

There has been a great revival of interest in one-dimensional
condensed matter systems in recent years. The technological
advances in the construction of quasi one-dimensional
hetero-structures have provoked a great deal of theoretical
research, investigating the transmission and reflection
coefficients
\cite{wave1,wave2,amp1,wave3,amp4,wave7,wave5,amp2,amp3,amp5,amp6,wave6,wave4,asr1,asr2,asr3,amplitp,ra1,amplitM},
energy band structures \cite{asr1,asr2,asr3}, and pattern of
resonant states \cite{asr1,asr2,asr3,ra1,r2} of one-dimensional
periodic systems. The effects of impurities and defects in
super-lattices have also been studied widely
\cite{impu0,impu1,impu2,impu10,impu3,impu4}. The impurities,
although unwanted in some cases, are found to be extremely useful
in some other cases like energy pass filters, quantum wires, and
waveguides \cite{impu1,impu2,impu3,impu10}. There have also been
studies on the total number of bound states and its relation to
the number of bound states of each constituting potential fragment
\cite{impu5,impu6,impu7,impu8,impu9}. A Levinson theorem has been
derived and used to study this relation in some works
\cite{impu5,impu6}.

The Levinson theorem for simple non-relativistic quantum
mechanical systems, in its original form \cite{Levi}, gave a
relationship between the $S$-state scattering phase shift
$\delta_0(k)$ and the total number of zero-angular-momentum bound
states $n_0$,
\begin{equation}
       \delta_0(0) -\delta_0(\infty) = \pi n_0 .   \label{eq:Leviorg}
\end{equation}
In 1957 Jauch \cite{Jauch} gave an alternative proof of
Eq.~(\ref{eq:Leviorg}) and generalized it to any angular momentum
state,
\begin{equation}
  \delta_l(0) -\delta_l(\infty) = \pi n_l .
\end{equation}
He also showed that the above relationship is a simple consequence
of the orthogonality and the completeness of the eigenfunctions of
the full Hamiltonian. In 1960 Newton \cite{Newton} showed that
when the potential is such that there exists a threshold bound
state with $l=0$, the statement of the Levinson theorem needs to
be modified to
\begin{equation}
  \delta_0(0) -\delta_0(\infty) = \pi \left( n_0 + \frac{1}{2}
  \right).
\end{equation}
A threshold bound state is most easily understood as a state with
zero momentum such that if the attractiveness of the potential is
increased infinitesimally, that state would emerge as a true bound
state. Much work in the literature has been devoted to the proof of
Levinson's theorem by different methods and its generalization to
scattering by non-spherically symmetric \cite{N2} or nonlocal
potentials (for a review see for example ref. \cite{rev} and the
references cited there). For an elegant new derivation see
\cite{Boya}.

In $1993$ a strong form of the Levinson theorem was presented in
the context of relativistic quantum mechanics in two space-time
dimensions  \cite {stronglev0,stronglev1}. The model considered consisted of
a Dirac particle coupled to a pseudo-scalar background field in a
solitonic configuration. The strong form of the Levinson theorem
related the phase shifts at each boundary of the continua $E=\{\pm
m, \pm \infty$ \} to the number of bound states that cross that
particular boundary, as the soliton evolves from the trivial
background. Moreover it was shown that the presence of the soliton
makes the phase shifts nontrivial at $E=\pm \infty$. The strong
form of the Levinson theorem for the finite boundaries of the
continua states that
\begin{equation}\label{levinsonstrong0}
\left. \delta(E)\right| _{E=\pm m} =(N_{exit}- N_{enter})\pi
\end{equation}
where $N_{exit}$ and $N_{enter}$ denote the number of bound states
that exit or enter the continuum from that edge, as the soliton or
any other disturbance is formed, respectively. Any threshold bound
state involved in this process counts as one half. That is,
$N_{exit}$ could involve a threshold bound state in two ways:
either a threshold bound state which turns into a complete bound
state, or a threshold bound state which appears at the band edge
originating from the continuum, as a distortion is formed.
Similarly $N_{enter}$ could involve a threshold bound state in two
ways which are exact opposites of the ones explained above. A
relationship analogous to Eq.(\ref{levinsonstrong0}) was also derived for
the infinite boundaries, which differs from it only by a minus
sign. If parity is a symmetry of the problem, the above statements
are true for each sign of parity separately. Combining these
statements, one can easily obtain the Levinson
theorem for the Dirac equation:
\begin{eqnarray}\label{levinsonweak0}
  \Delta\delta \equiv [\delta^{sky}(0) -\delta^{sky}(\infty)]+[\delta^{sea}(0) -\delta^{sea}(\infty)]  \nonumber\\
 = \pi \left( \mathcal{D}^{sky}+ \mathcal{D}^{sea} \right)=\pi \mathcal{D} =(N+\frac{N_t}{2}-\frac{N_{t}^{0}}{2})  \pi ,
\end{eqnarray}
where $\mathcal{D}$ denotes the total spectral deficiency, and
superscripts $sky$ or $sea$ denote the Dirac sky or sea,
respectively. $N$ is the total number of true bound states, $N_t$ is
the total number of threshold bound states at the given strength of
the potential, and $N_t^0$ is the total number of threshold bound
states at the zero strength of the potential, which is generically
non-zero in one-dimensional systems. The spectral density in both of
the continua, their position dependent deficiencies, and the local
and global completeness of the total spectrum in the presence of the
solitons were explicitly shown in \cite{nuclear}. We should mention
that there has been further works on the strong form of the Levinson
theorem, see for example \cite{stronglev2,stronglev3}. It is
interesting to note that, when the $E<0$ part is eliminated, both
forms of the Levinson theorem are also true in non-relativistic
quantum mechanics for isolated potentials, although the distinction
between the two forms is blurred by the fact that $\delta(\infty)$
can be usually set to zero in problems of physical interest.

In $1966$ Callaway \cite{Callaway} derived a form of the Levinson
theorem for the scattering of an excitation in a solid by a
potential of finite range,
\begin{equation}\label{Levwrong}
  \delta(E_s^l) -\delta(E_s^u) = \pi n_s,
\end{equation}
where $\delta(E_s^l)$ and $\delta(E_s^u)$ are the phase shifts at
the lower and upper energy edges of the $sth$ band, respectively,
and $n_s$ is the number of the states forced out of the band by the
perturbation. The quantity $n_s $ is exactly the same as the
spectral deficiency in the $s$th energy band, which can be denoted
by $\mathcal{D}_s$. The spectral deficiency can be simply defined as
the difference between the total number of states in the presence
and absence of the distortion. Due to recurring interest in
one-dimensional systems, this theorem has been derived and applied
to finite one-dimensional potentials where the half-bound states are
also taken into account
\cite{impu5,impu6,1dlev3,1dlev4,1dlev5,1dlev6,1dlev7,1dlev8}.
However, as we shall show, Eq.\ (\ref{Levwrong}), which is identical
to its counterparts in non-relativistic and relativistic quantum
mechanics for isolated potentials, is not always true and needs
several modifications. In this paper we obtain both forms of the
Levinson theorem for one-dimensional quasi-periodic potentials,
which, to the best of our knowledge, have not been presented before.
Moreover, we compare both forms to their analogues for isolated
potentials. Since this form of the Levinson theorem for
quasi-periodic cases is presented about 60 years after Levinson's
original derivation, we believe a simple illustration should
suffice. We hope to present its rigorous derivation later on.

As far as we know, the prevailing misconception is that the form of
the Levinson theorem for periodic potentials is exactly the same as
its form for isolated potentials.  Here we present a heuristic
derivation of the strong form of the Levinson theorem for
one-dimensional quasi-periodic potentials, which we believe could
clear up this misconception. In order to obtain the correct form of
the Levinson theorem for the solid state of matter, in particular
its strong form, we investigate the simplest non-trivial, yet
exactly solvable model, which is a distorted Kronig-Penney (KP)
model. This choice has the advantage of making the derivation of the
new forms of this theorem clear and simple. Moreover, although the
KP model \cite{krng1} is very simple and has been discussed in many
solid state textbooks, its generalizations have found wide usage in
investigating the essential features of more complex or
experimentally important structures such as super-lattices
\cite{krng2,krng3,krng4,krng5,krng6,krng7,krng8}. For this reason,
new methods are still being introduced to study the energy band
structure and eigenfunctions of KP models
\cite{krng9,krng10,krng11,krng12,krng13}. Therefore, this
investigation can also illuminate the essential features of the
Levinson theorem in more realistic models of solids.

To start the heuristic derivation we need to fully investigate the
physical properties of the undistorted system which is the simple
KP. However, since these are included in the standard textbooks we
shall do so very briefly, mainly to introduce our notation. In
Section two, we find the wave functions and energy band structure
for the simple KP model. However we discuss the parity eigenvalues
of states at the band edges in more detail, since the parity
assignments will be crucial for obtaining of the Levinson theorem.
We also discuss the criteria for the appearance of resonant states
within the bands and their effects on the parities of the states at
the band edges. In section three we calculate the bound states,
scattering states and their phase shifts for a distorted KP model.
The system has infinite spacial extent, and an infinite number of
bound states. However one can still define the phase shifts directly
from the scattering states (continuum eigenstates) of the full
Hamiltonian \cite{wang,sprung}. As we shall show, the phase shifts
have the correct limiting form in the sense that as the distortion
disappears the phase shifts go to zero. The phase shifts also
correctly count the bound states in the energy gaps and the spectral
deficiencies in the bands, exactly as one would expect from the
Levinson theorem. In this derivation we do not need to start with a
truncated form of the potential \cite{weber1,weber2}. Moreover it is
shown that there exists a strong form of the Levinson theorem which
relates the phase shifts at each band edge to the number of bound
states crossing that edge and the gap index, as the distortion is
formed. Then the relationship between the difference of the phase
shifts at the edges of a given energy band is related to the
spectral deficiency in that band and its index. These relationships
constitute modified forms of the Levinson theorem for the distorted
KP model. An additional relationship between the phase shifts at the
boundaries of any gap and the number of bound states in that gap is
also obtained.

\section{Essential Features of the Simple Kronig-Penney model}
\subsection{Eigenfunctions}
\begin{figure}[th]
\begin{center}\includegraphics[width=8.5cm]{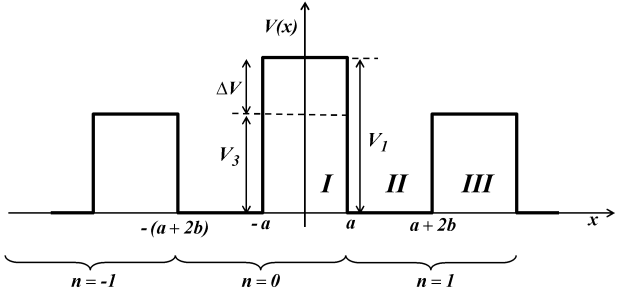}\caption{\label{fig1} \small
A distorted Kronig-Penney potential. Setting $\Delta V=0$, we
obtain a simple Kronig-Penney potential with lattice period
$l=2(a+b)$. $n=0,\pm1, ...$ denotes the cell number.}
\label{geometry}
\end{center}
\end{figure}

 The KP potential chosen is comprised of barriers of width $2a$
and height $V_3$, and wells of width $2b$. The lattice period is
thus $l=2(a+b)$, as shown in Fig.\ \ref{fig1}. For the simple KP
model $\Delta V$ shown in the figure is zero. The solution to the
Schr\"{o}dinger equation for this simple model is,
\begin{equation}
 \psi(x_n)=\left\{ \begin{array}{cc}
    A_ne^{ik_2(x_n-nl)}+B_ne^{-ik_2(x_n-nl)} \hspace{2mm}n\mbox{\small th well},\\
    C_ne^{ik_3(x_n-nl)}+D_ne^{-ik_3(x_n-nl)}  \hspace{2mm}n\mbox{\small th barrier},  \\
    \end{array} \right.
\end{equation}
where,
\begin{eqnarray}
n\mbox{\small th well} &:=& [(n-1)l+a] \leq x_n \leq nl-a,\\
n\mbox{\small th barrier} &:=& nl-a \leq x_n \leq nl+a,\\
k_2 &=& \sqrt{2\mu E/\hbar^2},\\
k_3 &=& \sqrt{2\mu (E-V_3) /\hbar^2},
\end{eqnarray}
and the subscript $n=0,\,\pm1,\,\pm2,\,...$ on $ A_n$, $B_n$, $
C_n$, $ D_n$ and $ x_n$ denotes the cell number. Using the
continuity of $\psi$ and its derivatives at well-barrier boundaries
of the nth cell, one can find the transfer matrix which relates the
coefficients in the adjacent wells, i.e. $A_n$ and $B_n$ to
$A_{n-1}$ and $B_{n-1}$ \cite{Merzbacher},
\begin{widetext}
\begin{equation}
 \begin{array}{c}
    \left(%
\begin{array}{c}
  {A_n} \\
  {B_n} \\
\end{array}%
\right)=
T\left(%
\begin{array}{c}
  A_{n-1} \\
  B_{n-1} \\
\end{array}%
\right), \\
\\
   T=\left(%
\begin{array}{cc}
  e^{2ik_2b}[\cos({2k_3a})+i\varepsilon/2\sin({2k_3a})] & i\eta/2\sin({2k_3a})e^{ik_2l} \\
  -i\eta/2\sin({2k_3a})e^{-ik_2l} & e^{-2ik_2b}[\cos({2k_3a})-i\varepsilon/2\sin({2k_3a})] \\
\end{array}%
\right), \\
 \end{array}
\end{equation}
\end{widetext} where $\varepsilon$ and $\eta$ are defined as
follows,
\begin{equation}
  \varepsilon \equiv \frac{k_2}{k_3}+\frac{k_3}{k_2}, \hspace{5mm} \eta \equiv \frac{k_2}{k_3}-\frac{k_3}{k_2}.
\end{equation}
The eigenvectors of the transfer matrix have the property,
\begin{equation}\label{e2:sec.term}
\frac{A_n}{B_n}=\alpha_{\pm}=\frac{i\eta\sin({2k_3a})e^{ik_2l}/2}{[\cos{(2k_3a)}+i\frac{\varepsilon}{2}\sin{(2k_3a)}]e^{2ik_2b}-t_\pm},
\end{equation}
where $t_\pm$ are the transfer matrix eigenvalues,
\begin{equation}
t_\pm=\frac{1}{2}[tr(T)\pm\sqrt{(tr(T))^2-4}].
\end{equation}
These satisfy the following important relationship,
\begin{equation}\label{dett}
\det(T)=t_+t_-=1.
\end{equation}
Therefore the wave functions in the $n$th cell can be written as,
\begin{equation}\label{e1:sec.term}
 \psi_{t\pm}(x_n)=t_\pm^n\left\{
\begin{array}{cc}
    A_0e^{ik_2(x_n-nl)}+B_0e^{-ik_2(x_n-nl)} \\
    C_0e^{ik_3(x_n-nl)}+D_0e^{-ik_3(x_n-nl)}
    \end{array} \right.
\end{equation}
For values of energy $E$ where $|tr(T)|\leq 2$, $t_\pm$ are
complex numbers of modulus unity:
\begin{equation}
t_\pm=e^{\pm ik_bl},
\end{equation}
and these define the allowed energy bands. For other energies
$t_{\pm}$ are real, therefore the solutions are divergent and these
define the band gaps. The special cases $t_{\pm}=1$ or $-1$, define
the band edges (see the discussion following Eq.\,(\ref{defp})).
Within the energy bands, Eq.~(\ref{e1:sec.term}) can be rewritten as
Bloch wave functions,
\begin{equation}\label{e7:sec.term}
 \psi_{t\pm}(x_n)=e^{\pm ik_bx_n}u_{k_b}^{\pm}(x_n),
\end{equation}
where $u_{k_b}^{\pm}(x_n)$ defined as,
\begin{equation}
u_{k_b}^{\pm}(x_n)\equiv e^{\mp ik_b(x_n-nl)}\left\{
\begin{array}{cc}
    A_0 e^{ik_2(x_n-nl)}+B_0e^{-ik_2(x_n-nl)}\\
    C_0 e^{ik_3(x_n-nl)}+D_0e^{-ik_3(x_n-nl)}
    \end{array}, \right.
    \end{equation}
are the cell periodic functions, which are invariant under any
lattice translation. It is obvious from Eq.(\ref{e7:sec.term})
that $k_b$ is the Bloch wave vector.

\subsection{Pattern of Parities at the Band Edges}

For the particular choice of the position of $x=0$ for setting the
symmetry point of the form of the potential, the parity symmetry is
manifest. Therefore, all the eigenstates of the Hamiltonian can be
chosen to have definite parities. The parity eigenstates can be
written as a linear combination of Bloch
 wave functions:
 \begin{equation}
 \psi_p(x_n)=A_{p,t+}\psi_{t+}+A_{p,t-}\psi_{t-},
 \end{equation}
where the subscript $p=\pm1$ denotes the parity eigenvalue. Writing
the wavefunction in the
 central barrier as,
 \begin{equation}
 \psi_p(x_0)=e^{ik_3x_0}+pe^{-ik_3x_0} \quad {-a \leq x_0 \leq a},
 \end{equation}
and matching the wavefunctions at the boundaries $x=\pm a$ the
expansion coefficients $A_{p,t\pm}$ are found to be,
 \begin{eqnarray}
   A_{p,t+}=\frac{A_0-p\alpha_{-}{A_0}^\ast}{\alpha_+-\alpha_-}, \nonumber\\
   A_{p,t-}=\frac{A_0-p\alpha_{+}{A_0}^\ast}{\alpha_--\alpha_+},
 \end{eqnarray}
where,
 \begin{equation}
A_0=\frac{e^{ik_2a}}{2}[(1+\frac{k_3}{k_2})e^{-ik_3a}+p(1-\frac{k_3}{k_2})e^{ik_3a}].
 \end{equation}

\begin{figure}[!]\begin{tabular}{c}
                    \includegraphics[width=7.5cm]{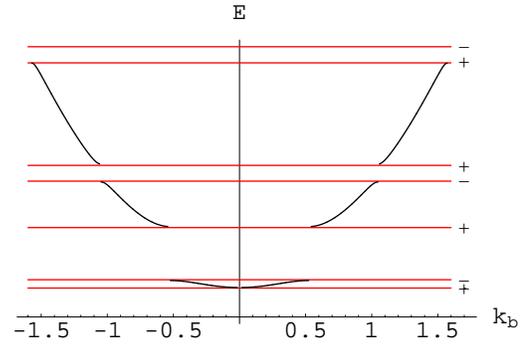} \\
                    \small{(a) Energy bands and parities at their
                    edges}\\ \\
                    \includegraphics[width=7.5cm]{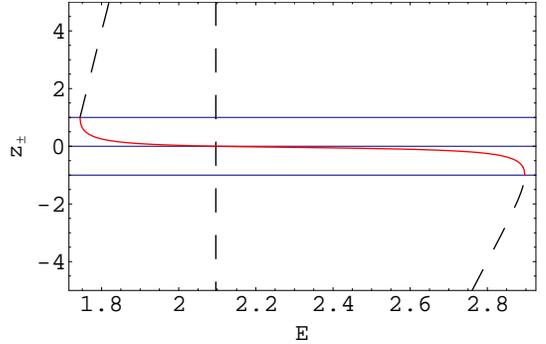} \\
                    \includegraphics[width=7.5cm]{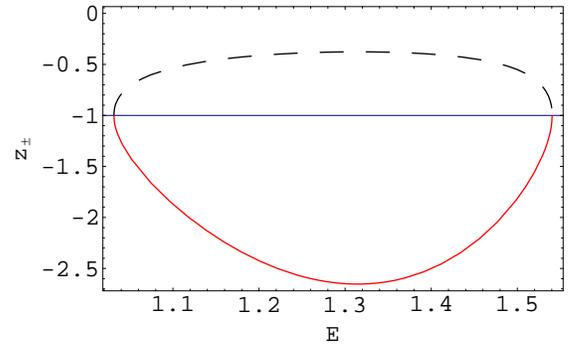} \\
                   \includegraphics[width=7.5cm]{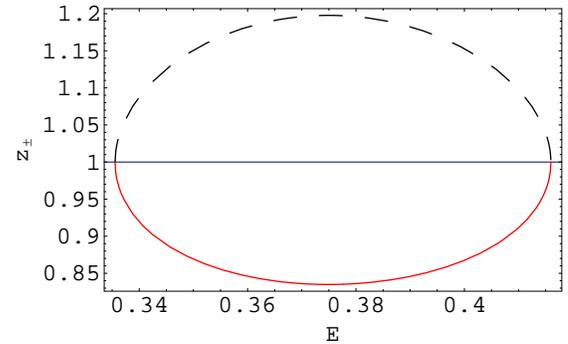} \\
                    \small{(b) $z_{\pm}$ in the first three energy band in ascending order}\\
                    \end{tabular}
\caption{\label{fig2} \footnotesize (a) The first three energy
bands of a KP potential with $a=b=1.5$ and $V_3=1$. The pattern of
parity signs for this potential is
\{$+,-,+,-,+,+,-,+,-,-,+,-,+,+,-,+,-,-,+,-,...$\}. Note the
periodic pattern which is particularly apparent if we disregard
the first two signs which belong to the zeroth band. The pattern
of periodicity depends on the parameters of the potential. (b) The
graphs of $z_{-}$ (solid lines) and $z_{+}$ (dashed lines) for the
first three energy bands. Note that these always have values $\pm
1$ at the band edges. Also note that in the third band $z_{\pm}$
change sign by passing through either zero or a divergent point,
which changes the sign of parities at {\em all} the higher band
edges. The particular energy in question is precisely where a
resonant state appears.} \label{geometry2}
\end{figure}

In the energy bands $\alpha_{\pm}$  can be written as,
\begin{equation}\label{alpha}
\alpha_{\pm}=z_{\pm}e^{ik_2l},
\end{equation}
where  the moduli $z_{\pm}$ are given by,
\begin{equation}
z_{\pm}=\frac{0.5\eta\sin(2k_3a)}{\cos(2k_3a)\sin(2k_2b)+\frac{\varepsilon}{2}\sin(2k_3a)\cos(2k_2b)\mp\sin(k_bl)}.
\end{equation}
Using Eqs.(\ref{alpha},\ref{e1:sec.term}) it can be shown that,
\begin{equation}\label{defp}
\psi_{t\pm}(-x)=t_{\pm}z_{\pm}\psi_{t\pm}(x).
\end{equation}

At the band edges $|tr(T)|=2$, and since Eq.(\ref{dett}) is true
everywhere, we have $t_+=t_-=1 \mbox{ or} -1$. Therefore,
Eqs.(\ref{e2:sec.term},\ref{alpha}) imply $z_+=z_-=1 \mbox{ or} -1$
at the band edges. Consequently the Bloch wave functions at these
values of energy become non-degenerate standing waves with definite
parity, whose signs are determined by the product of $t_{\pm}$ and
$z_{\pm}$, cf. Eq.(\ref{defp}). An alternative reasoning for
non-degeneracy of the states at the band edges is the following. At
the band edges the Bloch wave vector has modulus $n\pi/l$ and by
passing each barrier the wave function is only multiplied by $\pm1$,
consequently the degeneracy of the Bloch wave functions is broken at
the band edges and the eigenstates of the Hamiltonian are standing
waves with definite parities. These states can be considered as
threshold bound states in the absence of any distortion, as a
generic feature of any one-dimensional quantum mechanical system.
The wave function at the lowest allowed energy has always positive
parity. For a periodic potential consisting of delta-functions, the
sign of parity alternates at the band edges. However, as we shall
discuss, this alternating pattern changes as the barriers are
widened. The sign of $t_{\pm}$ at the edge of energy bands
alternates for all shapes of potentials but as the width of the
barriers grows from zero, the tendency of the wave functions to
avoid the barriers and gather in the wells makes $z_{\pm}$ change
sign in some energy bands. Therefore, Eq.\,(\ref{defp}) indicates
that the states at the top and bottom of such bands have the same
parity. This also leads to a corresponding change in the sign of
parities at all higher band edges. The eigenvectors of the transfer
matrix at energies where $z_{\pm}$ change sign by passing through
either zero or a divergent point, are pure left or right going plane
waves. These are called resonant states and are a result of the
reflection-less property of the potential at those energies. In
Fig.\ \ref{fig2}, $z_{\pm}$ and the parities at the band edges are
shown in the first three energy bands for a particular KP potential.
It is interesting that the parities at the band edges and the
appearance of resonant states follow some form of periodic pattern,
for any choice of the potential parameters, after skipping a few of
the lowest lying energy bands. The exact number of bands that needs
to be skipped before the periodic structure becomes manifest, and
the frequency of the appearance of the resonant states depends more
crucially on the width of the barriers. We make the following
labeling convention which shall be convenient when we formulate the
appropriate forms of the Levinson theorems: The very first gap, the
very first energy band, and the very first band edge are labeled as
zeroth gap, zeroth band and zeroth band edge, respectively.

\section{Distorted Kronig-Penney Model}

For simplicity we consider a particular distortion of the KP model
which is accomplished by only changing the strength of the central
barrier to $V_1=V_3+\Delta V$, as shown in Fig.\ \ref{fig1}. This
breaks the lattice translational symmetry; however, the Hamiltonian
is still invariant under the parity operator and the wavefunction in
the central barrier can be written as:
\begin{equation}
\psi_{d,p}(x_0)=e^{ik_1x_0}+pe^{-ik_1x_0} \quad {-a \leq x_0 \leq a},
\end{equation}
where the subscript $d$ denotes quantities in the distorted KP
model, and $p$ again denotes the parity eigenvalue and
\begin{equation}
k_1=\sqrt{\frac{2\mu}{\hbar^2}(E-V_1)}.
\end{equation}
From the boundary conditions at $x=\pm a$, the wave functions in
the zeroth and first wells are found to be
\begin{eqnarray}
\psi_{d,p}(x_0)&=&A_{d,0}e^{ik_2x_0}+B_{d,0}e^{-ik_2x_0},
 \nonumber\label{e3:sec.term}\\
& &\hspace{18mm}\mbox{\small in the zeroth well}\\
\psi_{d,p}(x_1)&=&A_{d,1}e^{ik_2(x_1-l)}+B_{d,1}e^{-ik_2(x_1-l)}\nonumber\\
& &\hspace{18mm}\mbox{\small in the first
well,}\label{e41:sec.term}
\end{eqnarray}
where
\begin{eqnarray}\label{e4:sec.term}
A_{d,1}&=&\frac{e^{ik_2(l-a)}}{2}[(1+\frac{k_1}{k_2})e^{ik_1a}+p(1-\frac{k_1}{k_2})e^{-ik_1a}]
\nonumber \\
 B_{d,1}&=&pA^*_{d,1},\,B_{d,0}=pA_{d,0}^*=pe^{-ik_2l}A_{d,1}.
\end{eqnarray}
Although the lattice translation symmetry is broken, we can still
generate the wave function in the nth well on the right or the left
by repeated application of $T$ or $T^{-1}$ on the wave function in
the first or the zeroth well, respectively. Therefore, the general
wave functions on the right and left sides of the central barrier
can be expanded in terms of the eigenstates of the transfer matrix
as follows
\begin{eqnarray}
\psi_{l}(x_n)=A_{l,t+}\psi_{t+}(x_n)+A_{l,t-}\psi_{t-}(x_n),
 \nonumber\\\label{e5:sec.term}
n\mbox{\small th well on the left}:\hspace{2mm}n=0,\,-1,\,\dots\\
\psi_{r}(x_n)=A_{r,t+}\psi_{t+}(x_n)+A_{r,t-}\psi_{t-}(x_n),
\nonumber\\\label{e6:sec.term}n\mbox{\small th well on the right}:
\hspace{2mm}n=1,\,2,\,\dots.
\end{eqnarray}

\begin{figure}\begin{tabular}{c}
   \includegraphics[width=8cm]{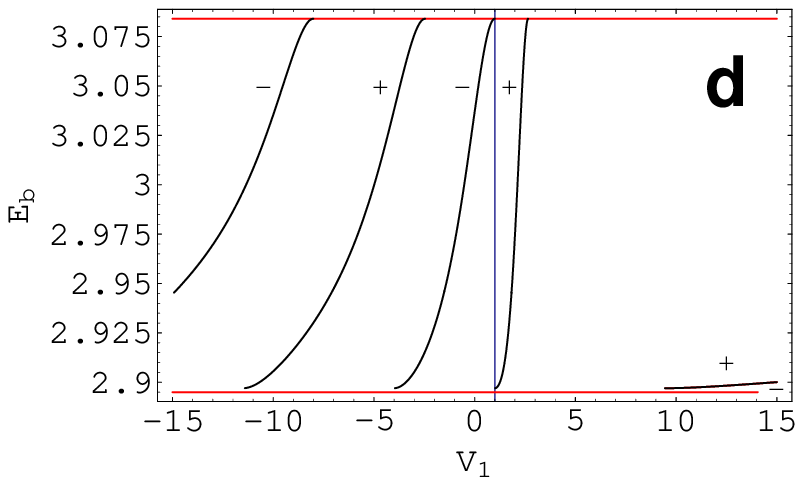} \\
  \includegraphics[width=8cm]{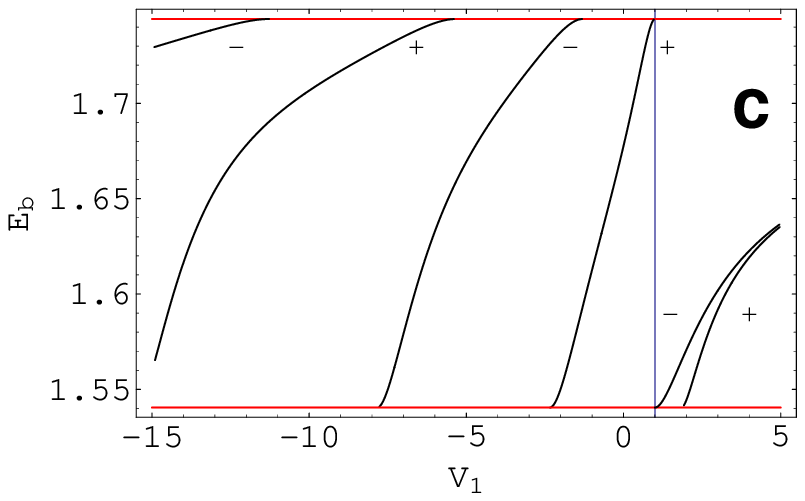} \\
  \includegraphics[width=8cm]{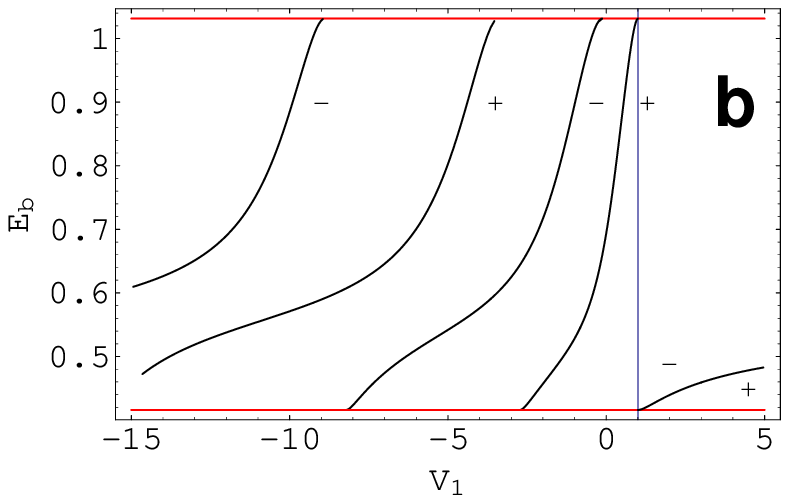} \\
   \includegraphics[width=8cm]{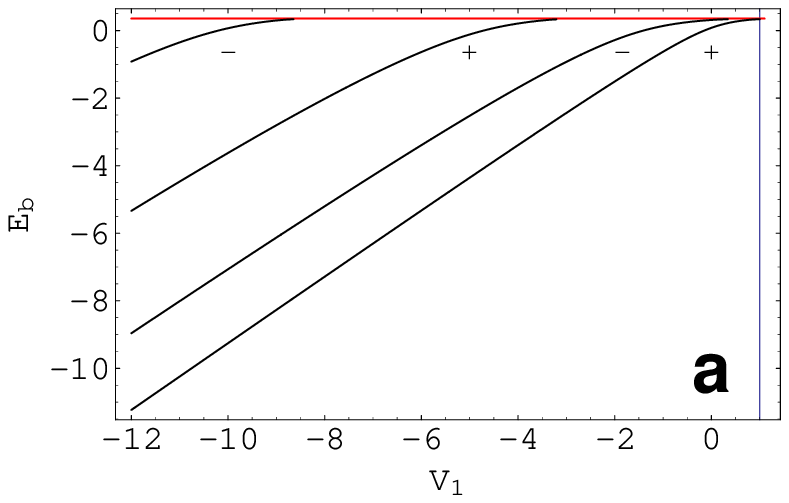} \\
\end{tabular}
\caption{\label{fig:4} \small The pattern of bound states which
appear due to the presence of a distortion ($\Delta V=V_1-V_3$), in
the first four gaps of a distorted KP potential versus the strength
of the central barrier $V_1$, depicted in ascending order(a,b,c,d).
The particular parameters of the potential are $a=b=1.5$ and
$V_3=1$. The case $V_1=1$ exactly corresponds to the simple KP
model. The two bound states for $V_1>1$ depicted in the first and
third gaps which seem to coincide on the scale shown, are actually
similar to the ones depicted in the second gap when exhibited in
smaller scale. The energies at the band edges are
\{0.3355,0.4160,1.0314,1.5405,1.7442,2.8970,3.0840,4.8727\}.}\label{geometry3}
\end{figure}

In general a limited local change in any periodic potential does
not change the overall structure of the continua, although one or
more states may be displaced below or above the continua, to form
localized or bound states in the forbidden energy zones. These
bound states appear at the cost of changes in the density of
states of the continua \cite{Callaway}. Since in the forbidden
gaps $t_{\pm}$ are real numbers, and Eq.~(\ref{dett}) is true
everywhere, one of the coefficients $A_{l,t+}$ or $A_{l,t-}$ in
Eqs.~(\ref{e5:sec.term}) must vanish for any normalizable solution
representing a bound state. A similar statement can be made for
the coefficients in Eqs.~(\ref{e6:sec.term}). Hence, due to the
parity symmetry, matching the two alternative solutions in the
first right well, for example, suffices for obtaining the
conditions for the occurrence of bound states:
\begin{equation}
\frac{A_{d,1}}{B_{d,1}}=\alpha_-\quad \mbox{if}\quad tr(T)>2,
\end{equation}
and
\begin{equation}
\frac{A_{d,1}}{B_{d,1}}=\alpha_+\quad \mbox{if}\quad tr(T)<-2.
\end{equation}
The energies of the bound states versus $V_1$ for the first four
energy gaps of a distorted KP potential are plotted in Fig.\
\ref{fig:4}.

\begin{figure}\begin{tabular}{cc}
     \includegraphics[width=4cm]{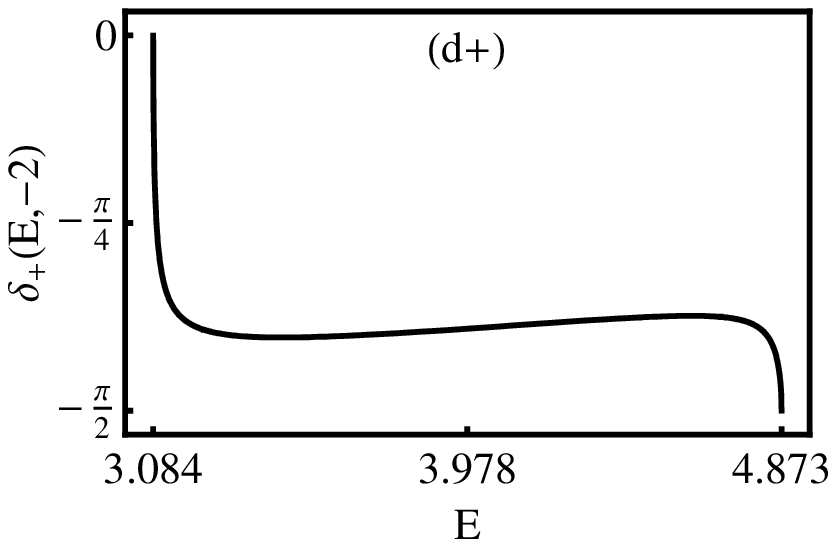}    \includegraphics[width=4cm]{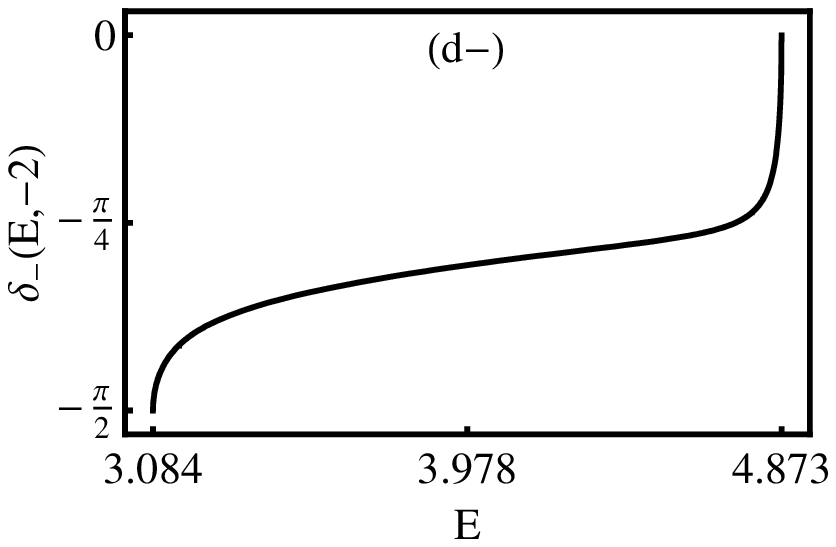}  \\
      \includegraphics[width=4cm]{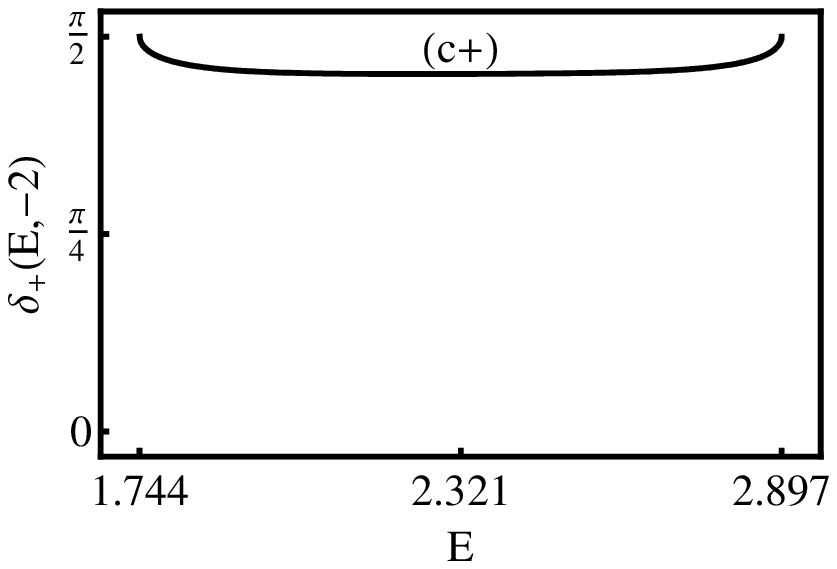}    \includegraphics[width=4cm]{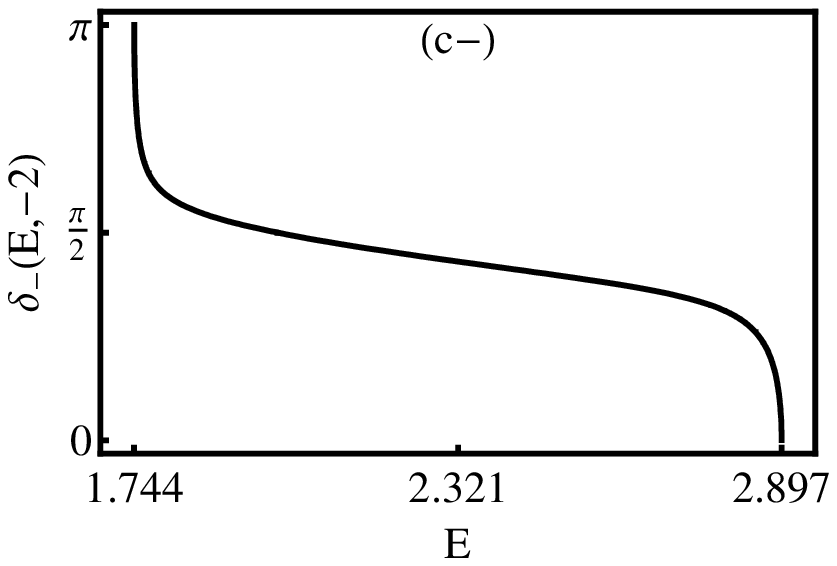}  \\
  \includegraphics[width=4cm]{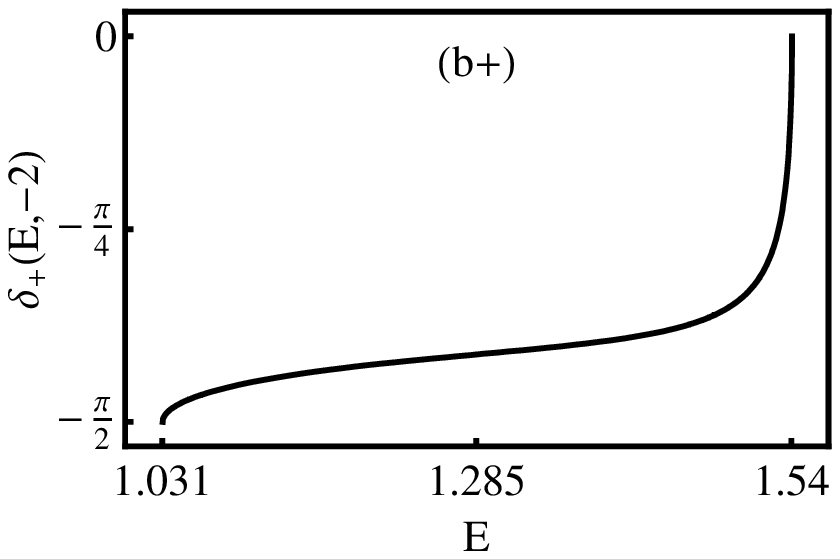}    \includegraphics[width=4cm]{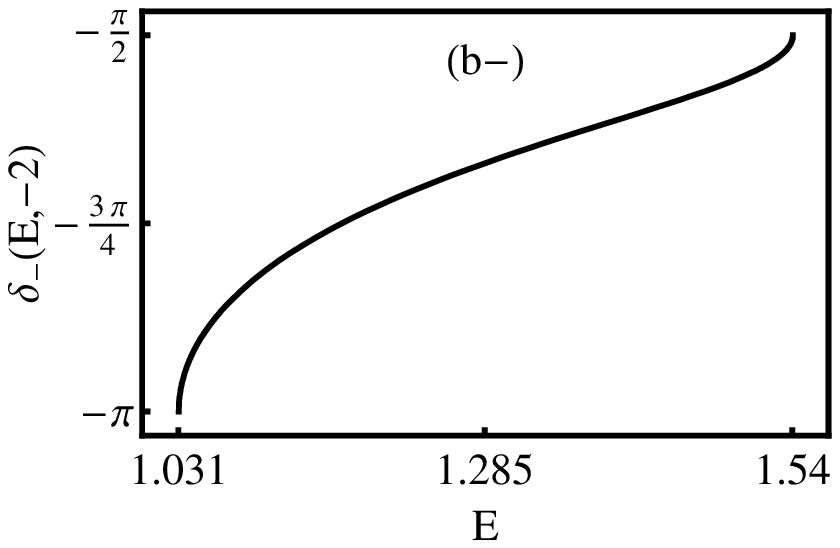}  \\
   \includegraphics[width=4cm]{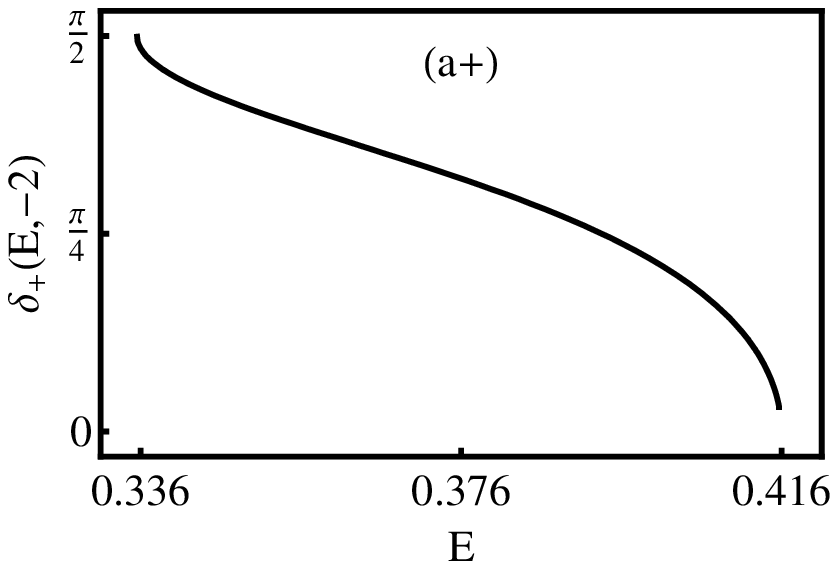}    \includegraphics[width=4cm]{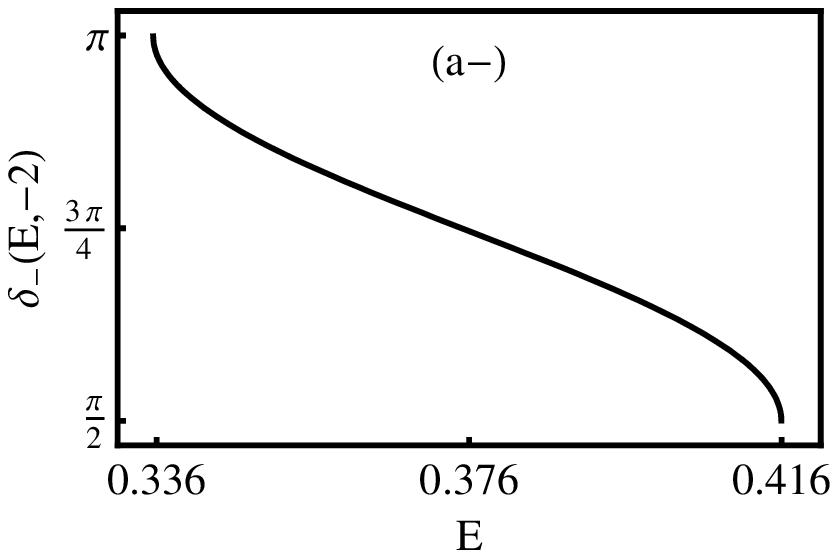}\\
\end{tabular}
\caption{\label{fig:1} \small The phase shifts in the first four
energy bands for the positive and negative parity eigenstates in a
distorted KP potential with $a=b=1.5,\,V_3=1$ and $V_1=-2$, depicted
in ascending order (a,b,c,d). Note that the phase shifts have values
which are integer multiple of $\pi/2$ at the band edges.}
\label{fig3}
\end{figure}

Now we can discuss the phase shifts within the allowed energy
bands. In the bands both $\psi_{t+}(x_n)$ and $\psi_{t-}(x_n)$ are
allowed and the expansion coefficients in Eqs.~(\ref{e5:sec.term})
and ~(\ref{e6:sec.term}) are found by matching the two alternative
solutions in the wells adjacent to the central barrier, as before,
\begin{equation}\begin{array}{ll}
A_{l,t+}=\frac{A_{d,0}-p\alpha_-A^*_{d,0}}{\alpha_+-\alpha_-},\quad A_{r,t+}=\frac{A_{d,1}-p\alpha_-A^*_{d,1}}{\alpha_+-\alpha_-}t_-
\end{array}
\end{equation}
\begin{equation}\begin{array}{ll}
A_{l,t-}=\frac{A_{d,0}-p\alpha_+A^*_{d,0}}{\alpha_--\alpha_+},\quad
A_{r,t-}=\frac{A_{d,1}-p\alpha_+A^*_{d,1}}{\alpha_--\alpha_+}t_+
\end{array}
\end{equation}
Substituting $\psi_{t+}(x_n)$ and $\psi_{t-}(x_n)$ from
Eq.~(\ref{e7:sec.term}) in Eqs.~(\ref{e5:sec.term}) and
~(\ref{e6:sec.term}), the scattering states in the bands can be
rewritten as:
\begin{eqnarray}
\psi_{p,l}(x_n)&=&A_{l,t+}u_{k_b}^+(x_n)e^{ik_bx_n}+A_{l,t-}u_{k_b}^-(x_n)e^{-ik_bx_n},
\nonumber\label{e8:sec.term}\\
& &\mbox{\small nth well on the left}:\hspace{2mm}n=0,\,-1,\,...\\
\psi_{p,r}(x_n)&=&A_{r,t+}u_{k_b}^+(x_n)e^{ik_bx_n}+A_{r,t-}u_{k_b}^-(x_n)e^{-ik_bx_n},
\nonumber\label{e8:sec.term}\\ & &\mbox{\small nth well on the
right}:\hspace{2mm}n=1,\,2,\,...
\end{eqnarray}
Now for each parity state the phase shift, $\delta_p(E,V_1)$, is
obtained by comparing the coefficients of $e^{ik_bx_n}$ on the
left and right-hand sides:
\begin{equation}
e^{2i\delta_p(E,V_1)}=\frac{A_{d,1}-p\alpha_-A^*_{d,1}}{A_{d,0}-p\alpha_{-}A_{d,0}^*}t_-.
\end{equation}
As the distortion is turned off, the bound states merge into the
bands and the phase shifts vanish, and this alleviates the inherent
ambiguity of the phase shifts by integer multiples of $\pi$. On the
band edges where $\alpha_+=\alpha_-$ it can be easily shown that,
\begin{equation}
e^{2i\delta_p(E,V_1)}=\pm1,
\end{equation}
which means the phase shifts are integer multiples of
$\frac{\pi}{2}$ at each band edge.

For simplicity, we consider a particular distortion and compute the
phase shifts of the parity eigenstates in the first four energy
bands of a distorted KP potential for $V_1=-2$ and plot the results
in Fig.\ \ref{fig3}. We base our conclusions on this set of
parameters. However, we have checked the validity of our forthcoming
conclusions for a wide range of parameters including changing the
strength of the central barrier and the widths $a$ and $b$.
Considering the historical background and the explanation given in
the introduction about the different contributions of half bound
states, one might expect that for a given band, the difference
between the phase shifts at the two band edges to count the number
of states which emerge out of that band minus the number that enter
that band, with the proper account of the threshold bound states.
Moreover, the strong form of the Levinson theorem for an isolated
potential states that the value of the phase shift at the edge of
the continuum is equal to the total number of bound states that have
exited minus the ones that have entered the continuum from that
edge, as the distortion is formed. However, as we shall see, these
are not quite true in the case of periodic potentials with isolated
distortions, and need one additional modification.

We now set out to extract the strong form of the Levinson theorem
from our results. Let us start with the zeroth energy band. As shown
in Fig.\ \ref{fig:4} (a), at zero strength of the distortion, i.e.
$V_1 - V_3 = \Delta V=0$, there is one positive parity threshold
bound state at the lower edge of the zeroth energy band which is
pulled down into the zeroth energy gap as $V_1$ is decreased.
Reducing the value of $V_1$ from $1$ to $-2$, a second bound state
with negative parity also emerges from this edge. Therefore, the
expected values for the phase shifts of positive and negative parity
eigenstates at this edge are $\frac{\pi}{2}$ and $\pi$,
respectively. The statement derived in \cite{stronglev0,stronglev1}
is that these phase shifts directly count the number of bound states
that emerge out of the band edge under consideration as the strength
of the potential is changed. For the aforementioned positive parity
bound state, it was already a threshold state at zero distortion and
its emergence as a complete bound states counts as one half. These
values are in agreement with phase shift diagrams in Fig.\
\ref{fig3}(a). At the upper edge of the zeroth energy band, Fig.\
\ref{fig:4} (b), there is a negative parity threshold bound state at
zero distortion, which sinks into the band as $V_1$ is reduced to
$-2$. Thus the phase shifts at this edge are expected to be zero and
$-\frac{\pi}{2}$ for positive and negative parity eigenstates,
respectively. However, the phase shift values at this edge, as shown
in Fig. \ref{fig3} (a), differ by a minus sign from what is
speculated.

At the lower edge of the first energy band, there is a positive
parity threshold bound state at zero strength of the distortion,
which is pulled down to the first energy gap as the value of $V_1$
is reduced from $1$, Fig.\ref{fig:4} (b). Later a negative parity
bound state emerges from this edge when $V_1$ is reduced further
towards $-2$. The expected phase shift values for positive and
negative eigenstates at this edge are $\frac{\pi}{2}$ and $\pi$,
respectively, which again differ by a minus sign from the actual
phase shift values at this edge, as shown in Fig.\ref{fig3} (b).

Comparison between the phase shifts of eigenstates in higher
energy bands and the number of bound states displaced out of or
into these bands mandate a different rule for the strong form of
the Levinson theorem for periodic potentials, which is
\begin{equation}\label{levinsonstrong}
\delta_{\pm}(E_s,V_1)=(-1)^s(N_{exit}^{s,\pm}- N_{enter}^{s,\pm})\pi,
\end{equation}
where $s$ is the energy \emph{gap} index, $E_s$ labels the energy
value at either the lower or upper boundary of the $s$th gap, and
$N_{exit}^{s,\pm}$ and $N_{enter}^{s,\pm}$ are the total number of
bound states of definite parity that exit or enter the energy bands
from that particular boundary. Any threshold bound state involved in
this process counts as one half, as explained in the introduction,
cf. Eqs.(\ref{levinsonstrong0},\ref{levinsonweak0}) and their
following explanations. Subtracting the expression for the phase
shifts at the lower and upper edges of the $s$th energy band, the
Levinson theorem is easily obtained,
\begin{eqnarray}\label{levinsonweak}
  \Delta\delta_{s,\pm} & \equiv & \delta_{\pm}(E_s^l,V_1) -\delta_{\pm}(E_s^u,V_1)  \nonumber\\
  &=&\pi(-1)^s(N_{out}^{s,\pm}-N_{in}^{s,\pm}) =\pi (-1)^s \mathcal{D}_{s,\pm},
\end{eqnarray}
where $s$ is the energy \emph{band} index, $\delta_{\pm}(E_s^l,V_1)$
and $\delta_{\pm}(E_s^u,V_1)$ are the phase shifts of eigenstates of
definite parity at the lower and upper boundaries of the $s$th
energy band. $N_{out}^{\pm ,s}$ and $N_{in}^{s,\pm}$ denote the
number of bound states that leave or enter the $s$th band as the
distortion is turned on. Threshold bound states count as one half,
exactly as explained before. The quantity $\mathcal{D}_{s,\pm}$
denotes the spectral deficiency in the $s$th band, for each parity
separately. This formula can be easily cast into the form of
Eq.~(\ref{levinsonweak0}). Therefore, the only difference between
both forms of the Levinson theorem for the distorted KP potential
and their forms for isolated potentials in relativistic
\cite{stronglev0,stronglev1} or non-relativistic quantum mechanics
is the factor $(-1)^s$. By adding the phase shifts at the lower and
upper boundaries of the $s$th gap, we can also present the following
interesting relationship between that quantity and the number of
bound states present in that gap due to the distortion, $N_{s,\pm}$,
\begin{equation}\label{boundstatesingap}
\delta_{\pm}(E_s^l,V_1) +\delta_{\pm}(E_s^u,V_1)=\pi(-1)^s N_{s,\pm}.
\end{equation}
We can combine Eqs.~(\ref{levinsonweak},\ref{boundstatesingap}) to
relate the total number of bound states, deficiencies, and the phase
shifts
\begin{eqnarray}\label{total}
\delta_{\pm}(E_0,V_1) &-&\delta_{\pm}(E_{1},V_1)-\delta_{\pm}(E_2,V_1)+  \\
\delta_{\pm}(E_{3},V_1)&+&\delta_{\pm}(E_4,V_1)+ \dots=\pi\Sigma N_{s,\pm}=\pi\Sigma \mathcal{D}_{s,\pm},\nonumber
\end{eqnarray}
where the energies of the boundaries are labeled consecutively for
clarity. This equation makes the completeness relationship manifest.
For the special case of the zeroth gap and band edge
Eq.~(\ref{boundstatesingap}) reduces to,
\begin{equation}\label{zero}
\delta_{\pm}(E_0,V_1))= \pi N_{0,\pm},
\end{equation}
which is what we expect from Eq.~(\ref{levinsonstrong}).

\section{Conclusions}

In this paper we calculated the band structure, continuum states and
their phase shifts, and the bound states for a distorted KP model.
We obtain the strong form of the Levinson theorem as stated in
Eq.(\ref {levinsonstrong}) which relates the phase shifts at each
band edge to the number of states that cross that edge. Obviously
threshold bound states count as one half as explained in the text.
From this theorem we can easily conclude the Levinson theorem as
stated in Eq.(\ref {levinsonweak}), which relates the difference
between the phase shifts at edges of a given band to the spectral
deficiency of that band. Both theorems are identical to their
counterparts for localized potentials in relativistic and
non-relativistic quantum mechanics, except for a factor of $(-1)^s$,
where $s$ denotes the adjacent gap or the band index, for the strong
or weak form of the theorem, respectively. We have also obtained a
relationship between the phase shifts at the edges of any gap to the
number of bound states present in that gap Eq.(\ref
{boundstatesingap}), and an overall relationship exhibiting
completeness of the spectrum Eq.(\ref {total}).

\vskip20pt\noindent {\large {\bf
Acknowledgements}}\vskip5pt\noindent

We would like to thank the research office of the Shahid Beheshti
University for financial support.

\end{document}